# The Interstellar Medium White Paper
*A discussion paper for Australian engagement within this arena of astronomy*

Michael Burton (University of New South Wales), Roland Crocker (Australian National University), John Dickey (University of Tasmania), Miroslav Filipovic (University of Western Sydney), Cormac Purcell (University of Sydney), Jill Rathborne (CSIRO Astronomy and Space Science), Gavin Rowell (University of Adelaide), Nick Tothill (University of Western Sydney), Andrew Walsh (Curtin University).

*Version 2, 15 December 2013*

**Abstract**
The interstellar medium is the engine room for galactic evolution. While much is known about the conditions within the ISM, many important areas regarding the formation and evolution of the various phases of the ISM leading to star formation, and its role in important astrophysical processes, remain to be explained. This paper discusses several of the fundamental science problems, placing them in context with current activities and capabilities, as well as the future capabilities that are needed to progress them in the decade ahead. Australia has a vibrant research community working on the interstellar medium. This discussion gives particular emphasis to Australian involvement in furthering their work, as part of the wider international endeavour.

The particular science programs that are outlined in this White Paper include the formation of molecular clouds, the ISM of the Galactic nucleus, the origin of gamma-rays and cosmic rays, high mass star and cluster formation, the dense molecular medium, galaxy evolution and the diffuse atomic medium, supernova remnants, the role of magnetism and turbulence in the Galactic ecology and complex organic molecules in space.

# Overview

The interstellar medium (ISM) is a dynamic, ever-changing environment. It provides the cauldron from which stars are born. It serves as the medium for receiving the enriched products of stellar nucleosynthesis expelled from stars during their evolution. The continual energy flows through the ISM and the interchange of materials between its gas and that in stars forms a cycle that drives the evolution of galaxies. The ISM encompasses ionized, neutral and molecular environments with density and temperature contrasts of many orders of magnitude; roughly $10^{-3} – 10^{10}$ cm$^{-3}$ and $10 – 10^7$ K, respectively. It also contains a solid component – dust grains that can provide the seeds for planet formation. While the basic parameters of the ISM are known – the range of environments that exist and their contributions to the mass and volume budgets of our Galaxy, for instance – many mysteries remain. The formation processes of clouds, both atomic and molecular, are unclear, the key events which determine when and how star formation occurs are contentious, the ISM's role in the some of the highest energy astrophysical phenomena is still unknown, and the level of complexity of the organic molecules it can produce still undetermined.

Active research programs are taking place in Australia across a wide range of ISM studies. There is a vibrant research community working within this arena. Australian facilities have contributed greatly to our knowledge of many aspects of the ISM, as well as playing important roles as part of wider international investigations. This White Paper provides a look forward to the decade ahead, outlining science programs for several fundamental and challenging problems relating to the evolution of the ISM and the formation of stars within it. These are science problems where Australians have already played leading roles in the international endeavour, and can continue to do so in the years ahead given appropriate investment in resources. In this document several representative problems are outlined, the capabilities needed to progress them described, and the role of Australian and international facilities – both current, or with enhanced capabilities, or entirely new – explained. The Science programs discussed in the pages ahead include:

- The formation of molecular clouds
- The interstellar medium of the Galactic nucleus
- The origin of cosmic rays and the production of gamma rays
- High mass star and cluster formation
- Charting the dense molecular medium of the Galaxy
- Galaxy evolution and the diffuse atomic gas – GASKAP and HI
- Supernova remnants
- The role of magnetic fields and turbulence in the Galactic ecology
- The environments for complex organic molecules in space

Capabilities across the infrared to radio wavebands are essential for these endeavours, but particularly so for the radio regime, in the centimetre, millimetre and terahertz bands. Central to the furthering the research goals are the capability to undertake wide field spectral line surveys of the ISM across a range of frequency bands, together with the ability to undertake targeted high spatial resolution follow-up of sources of particular interest. The facility suite includes access to the grand design interferometers of the SKA and ALMA, as well as the ATCA, MWA and ASKAP, together with single dish radio telescopes such as Parkes, Mopra, Nanten2, APEX and Antarctic THz facilities (HEAT, STO and DATE5). In high energy gamma-ray astrophysics the progression from HESS to the Cherenkov Telescope Array (CTA) demands complementary capability across the radio bands for large areal surveys on the arc-minute scale. Equally important is a science community with strengths across observational, instrumental and theoretical areas. The ability to use all these facilities to their full capability requires scientists who can contribute to our understanding of the evolution of the ISM in its many forms and environments.

# Wide Field Surveys with Targeted Follow-Up
*An Overview of the Facility Needs to Support the Australian Community*

| Principal Species | Freq. GHz | Current Facilities | | Future Facilities |
|---|---|---|---|---|
| H | 1.42 | 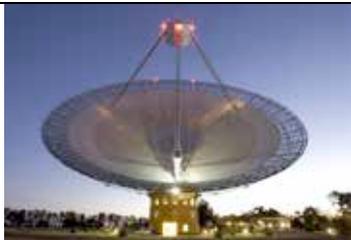 Parkes 64m | 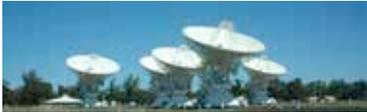 ATCA 6 x 22m | 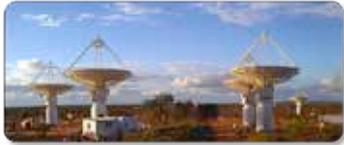 ASKAP 36 x 12m |
| CO | 110, 115 | 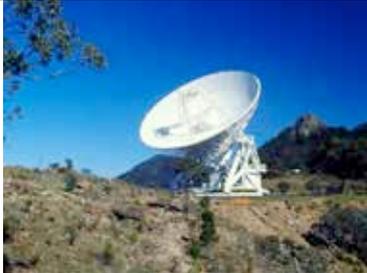 Mopra 22m | 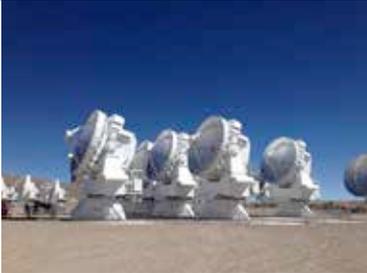 ALMA 66 x 12m, Chile | 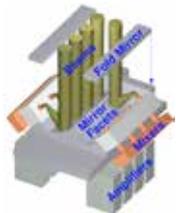 Mopra Multibeam |
| C | 491, 809 | 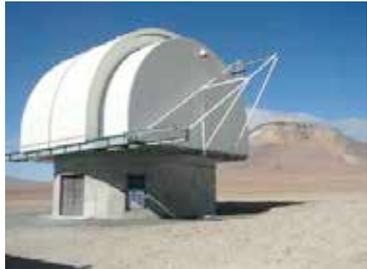 Nanten2 4m, Chile | 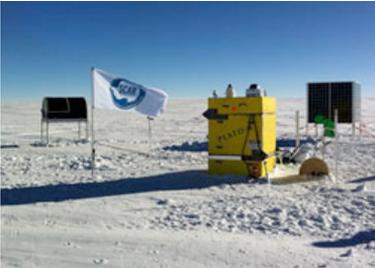 HEAT 0.6m, Ridge A | 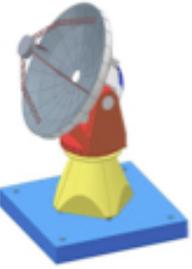 DATE5 5m, Dome A |
| $C^+$ | 1900 | | | 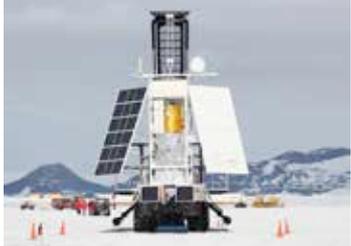 STO 0.8m, McMurdo |

# The Formation of Molecular Clouds
*Following the Galactic Carbon Trail*

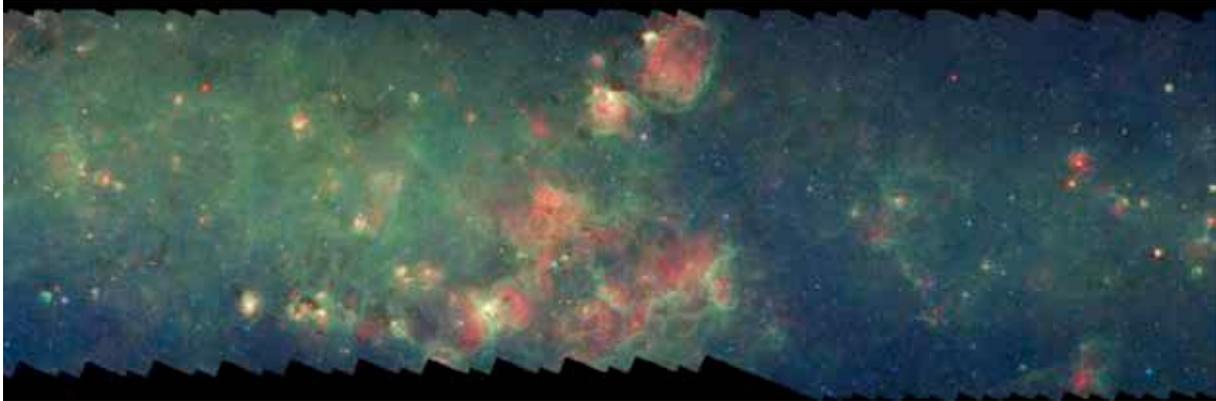

*A 7° portion of the southern Galactic plane seen in the infrared from the Spitzer space telescope, showing sites of active star formation. Red generally represents warm dust emission at 24μm where star formation is active, green fluoresced PAH emission at 8μm and stars are seen in blue at 4μm. Molecular clouds must also be continually forming within this region before star formation can commence; see* www.phys.unsw.edu.au/mopraco. *Image courtesy of NASA/JPL-Caltech.*

*Overview*
Stars are born within molecular clouds, a process that begins almost as soon as the natal molecular gas has itself formed. But how do molecular clouds form in the first place? This fundamental process has yet to be observed. It is the rate-determining step for star formation itself.

Observations of external spiral galaxies show that massive stars and their giant molecular clouds (GMCs) tend to form in the compressed regions of spiral arms, behind the spiral density wave shock (Engargiola et al. 2003). If these regions are primarily atomic, then the atomic gas must somehow collected together to form GMCs. If instead they are mainly molecular, such as in our Galaxy's molecular ring, then the collection into GMCs may involve small molecular clouds bound together by pressure rather than self-gravity. Several mechanisms have been proposed for this process of GMC formation. They include (see Elmegreen 1996):
(i)     self gravitational collapse of an ensemble of small clouds,
(ii)    random collisional agglomeration of small clouds,
(iii)   accumulation of material in shells, e.g. as driven by winds and supernovae, and
(iv)    compression and coalescence of gas in the converging flows of a turbulent medium.

These scenarios provide quite different pictures for the structure and evolution of GMCs. For instance, under mechanism (i) the gravitational collapse of a cluster of clouds produces molecular clouds that are long-lived and stable, supported against gravity by internal turbulence and magnetic fields. This contrasts strongly with the picture given by mechanism (iv), of compression in converging flows, where gravity plays little role. It produces molecular clouds that are transient features. Furthermore, more than one of these mechanisms may be at work.

*Morphological and Kinematic Signatures of Molecular Cloud Formation*
In turn, each of these scenarios provides different observational signatures that will allow them to be distinguished. For instance, if clouds form by the gravitational collapse of a cluster of small clouds [i], there should be a roughly spherical, or possibly filamentary, distribution of small clouds. If molecular clouds form by random collisional coagulation of small clouds [ii], the velocity field of the cluster clouds will look more random and less systematic than infall. If they are formed in wind or supernova-driven shells [iii], the shell-like morphology will be apparent. If they are formed by converging flows in a turbulent interstellar medium [iv], then there should be an overall a turbulent velocity field, but local to the formation sites velocities will be coherent (i.e. converging).

*Capability Requirements to Observe the Signatures*
In order to contain sufficient mass to build a GMC, an atomic cloud needs to have a hydrogen column of order $10^{21}$ cm$^{-2}$ so as to shield any molecules that form from dissociating UV radiation. This corresponds to a diameter of about 7pc at interstellar pressures. Small molecular clouds also need similar columns, but are cooler and denser than atomic clouds and so may have sizes of order 1–2pc. GMCs themselves have diameters of 10–100pc. Ensembles of small clouds that may coalesce into GMCs will have diameters of 200–1000pc. A survey undertaken to search for such objects in our Galaxy will pass through the molecular ring at distances of typically 4–8kpc. Therefore these sizes correspond to 0.5'–2' for small clouds, 4'–80' for GMCs and 1°–10° for the ensembles. This implies a capability requirement of 0.5' spatial resolution to resolve individual clouds and a survey area that extends across a spiral arm and is at least three times larger than the ensemble size; i.e. >30° to ensure that it encompasses the complete range of phenomena occurring.

Linewidths observed toward small individual clouds are of order 1 km/s and toward GMCs ~5 km/s. Velocity information is generally used to place the clouds along the line of sight, using the galactic rotation curve. Typically 1 km/s corresponds to ~100pc in distance. However, if clusters of clouds are seen in the two dimensions on the sky, then it is possible to also determine their velocity dispersion by eliminating any spatial elongation (akin to the "finger of God" structures seen in galaxy redshift surveys) along the sight lines. Spectral resolutions of ~1 km/s are therefore needed to determine the overall 3D distribution of cloud clusters and ~0.1 km/s for the velocity distributions of infalling clouds within these clusters.

*Diagnostic Tracers – Carbon not Hydrogen*
Hydrogen is the dominant constituent in gas clouds, found as H and $H_2$ in the atomic and molecular clouds. However, while hydrogen can readily be detected through its 21cm HI line, this does not provide a ready distinction as to whether the emitting gas comes from the cold neutral or warm neutral phases (the former being the precursor to molecular clouds), or provide the physical state of the gas (i.e. density and temperature). Moreover, the lines are also broad, making it difficult to separate narrow components within them. Molecular hydrogen, however, cannot even be observed from cold molecular clouds. Its lowest energy transition arises from 500K above ground and so is not excited in the typical 10–20K temperatures found in molecular gas.

Carbon, with an abundance of ~$10^{-4}$ of hydrogen, cannot be hidden, however. It will exist in ionized ($C^+$), neutral (C) or molecular (CO) form, dependent on the physical state of the gas. Deep within molecular clouds it can be followed through millimetre CO line emission at 115 GHz. Close to the atomic/molecular interface carbon will neutral, emitting through the sub-millimetre lines of [CI] at 492 and 809 GHz. Across the interface region into the atomic medium it will emit in the terahertz band, through the [CII] line at 1.9 THz.

*Facilities Required – Following the Galactic Carbon Trail*
Thus wide-field, moderate spatial and high spectral resolution survey capability is needed across the THz to millimetre-wave bands. The 22m Mopra telescope provides the needed capability for CO, although the single pixel detector limits survey areas to about 10 sq. deg in a season – a multibeam detector is desirable (Burton et al. 2013a). The 4m Nanten2 telescope in Chile, equipped with a 8-pixel array, provides a corresponding capability at 492 GHz for [CI]. The 60cm HEAT telescope at Ridge A in Antarctica is currently demonstrating the capability to survey for [CI] at 809 GHz, though is limited to a 2' beam (Kulesa et al. 2013, Burton et al. 2013b); this will be much improved by the Chinese-led 5m DATE5 telescope for Dome A. When the water vapour column drops below 100μm at Ridge A it is anticipated that [CII] measurements will also be possible there. Alternatively, balloon-borne telescopes, such as STO and GUSSTO, proposed for long-duration balloon programs from McMurdo station (Walker 2012), provide the needed capability.

# The ISM of the Galactic Nucleus
*From the Central Super-Massive Black Hole to Galactic-Scale Radio Lobes*

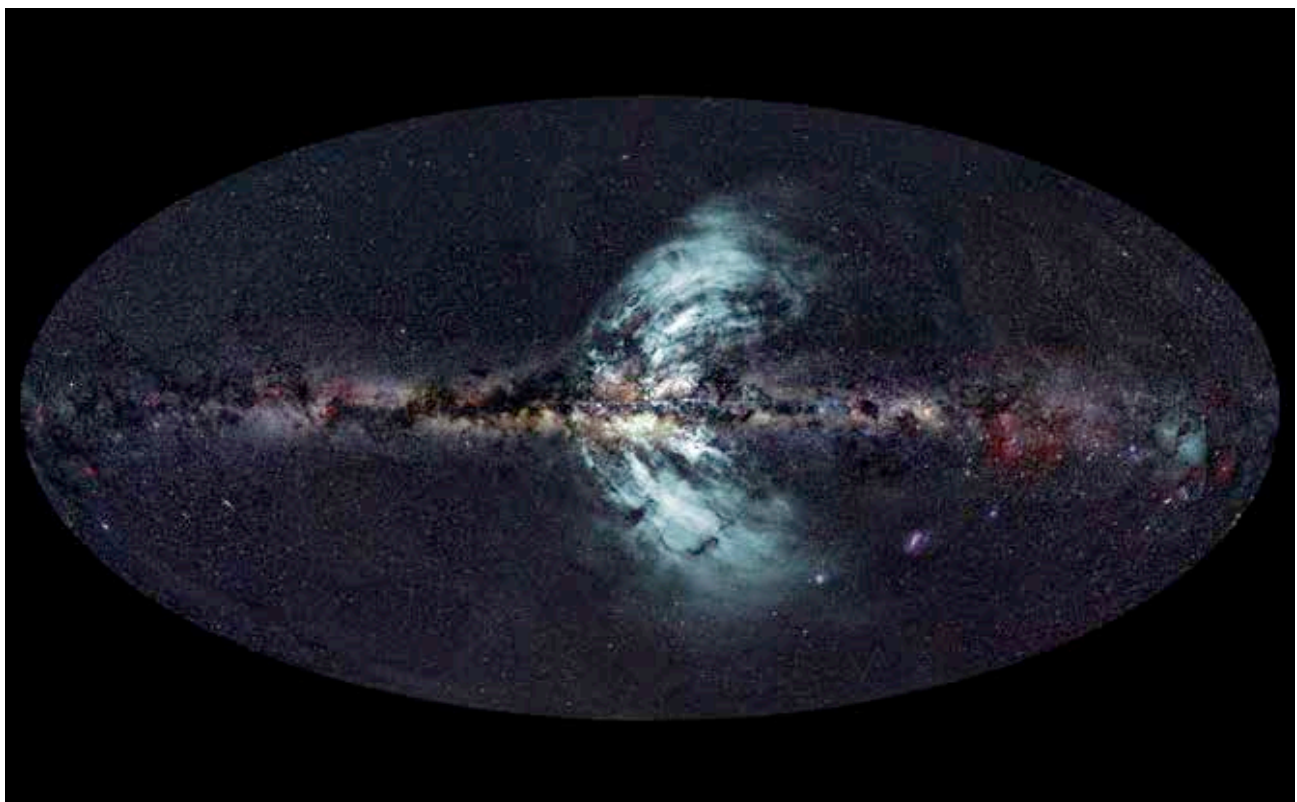

*The giant radio lobes emanating from the Galactic Centre (pale blue) recently discovered with the Parkes telescope. The background image is the whole Milky Way at the same scale. Credits: Ettore Carretti, CSIRO (radio image); S-PASS survey team (radio data); Axel Mellinger, Central Michigan University (optical image); Eli Bressert, CSIRO (composition).*

The central few hundred parsecs of the Galaxy represent an environment quite different to the rest of the Galaxy (Morris and Serabyn 1996). Because gas is always falling inwards towards the Milky Way's central regions, the Galactic nucleus has, over long periods, sustained an intensity of star-formation much higher than the (on average) more quiescent Galactic disk (Serabyn and Morris 1996). The radiation, winds, and supernova explosions that result from the ongoing star formation in the nucleus heat, energize and over-pressure the nuclear environment. This, in turn, creates giant outflows out of the central few hundred parsecs that loft plasma, cosmic rays, and entrained magnetic fields high up into the Galactic halo (Crocker et al. 2011a, Crocker et al. 2011b, Crocker 2012). These outflows have been detected in gamma-rays (as the 'Fermi Bubbles'; Su et al. 2010) and, most recently, as giant, highly-magnetized radio lobes. These lobes, discovered only last year by Ettore Carretti and colleagues using the Parkes radio telescopes, constitute the second largest structures in the Galaxy: with radio eyes, one would seem them crossing more than half the sky (Carretti et al. 2013).

It is surprising indeed that such large structures as these Lobes have remained undiscovered until so recently. From an observational perspective this highlights the need for further sensitive and large angular scale radio polarimetry surveys to trace outflows from both the nucleus of our own and from external galaxies. Low frequency SKA pathfinders including MWA will contribute significantly in this endeavour. The discovery of the Fermi Bubbles in data supplied by the orbiting Fermi telescope highlights the growing importance of gamma-ray data in astrophysics. It is crucial that Australian astrophysics and astronomy grows its connection to this rapidly-developing field, in particular, via a strong involvement with the Cherenkov Telescope Array (CTA) project being led from Europe.

From a theoretical perspective, although the picture is becoming steadily clearer, there are many aspects of the dynamics of the Galactic nucleus and its outflow that we do not understand. It is still under debate, for instance, whether it has indeed been star formation (Crocker and Aharonian 2011) or, alternatively, activity of the central super-massive black hole (e.g. Su & Finkbeiner 2012) that was primarily responsible for the inflation of the Fermi Bubbles and the corresponding radio lobes. We also cannot yet be sure whether the Bubbles are quasi steady state phenomena or the signatures of much more recent and transient events.

Such questions feed into one of the most important problems in modern astrophysics: to understand where super-massive black holes (SMBH) come from and how their presence influences their host galaxy. In spiral galaxies, the SMBH is not the only important actor in the nucleus, however. Statistical studies show that there are scaling relations connecting not only the SMBH but also the nuclear stellar population to galaxy-wide properties of the host (Launhardt et al. 2002). These relations signal the activity of 'feedback' processes wherein activity in a galactic nucleus is communicated out to large scales by giant outflows. Due to its proximity, the centre of the Milky Way is one of the best environments to study these feedback processes in action and the Fermi Bubbles seem to present direct evidence of such feedback processes in action.

It must finally be mentioned that – even setting aside all the spectacular evidence for huge outflows – an understanding of star-formation in the Galactic nucleus is a crucial prerequisite to understanding the Galaxy's overall star formation ecology. This tiny region is responsible for at least 10% of the *massive* star formation occurring in the Milky Way. Furthermore, to understand the star formation occurring there we must understand the dynamics of the molecular gas that is the fundamental fuel from which new stars are grown. This points to the importance of ongoing efforts with the Mopra (see [www.phys.unsw.edu.au/mopracmz](www.phys.unsw.edu.au/mopracmz)), Nanten2 and other telescopes to collect deeper and higher resolution molecular line data revealing details of the inflow and outflow of such gas as well as the dense molecular gas cores where star formation is actually occurring. A deeper understanding of star-formation in the Galactic centre context also holds out the promise of elucidating star formation processes occurring in dense star-burst environments in both the local and high redshift universe for which the Galactic centre provides the closest analogue environment.

# The Origin of Cosmic Rays and the Production of Gamma rays

*Overview and Key Issues*

The distribution and dynamics of the ISM plays a pivotal role in understanding the nature of gamma-ray sources as revealed in recent years by instruments such as Fermi-LAT at GeV energies, and HESS, MAGIC and VERITAS at TeV energies respectively. The current instruments utilise techniques that have matured over the past 40 years. In this overview we focus on aspects related to TeV gamma ray energies, drawing on recent reviews and selected results (Aharonian et al 1991, Amato et al. 2003, Greiner et al. 2005, Aharonian et al. 2005, 2006a, Hinton & Hofmann 2009, Actis et al. 2011, Nolan et al 2012, Ackermann et al. 2013, Rieger et al. 2013, Acero et al. 2013).

Gamma-ray astronomy is tightly connected to the origin of cosmic-rays (CRs) and ISM properties. ISM gas is a target for CR collisions, which lead to >GeV gamma-ray emission. Gamma-rays therefore trace regions of CR interaction and acceleration. For CR collisions, the ISM can be in *any* chemical state and so gamma-ray emission provides an estimate of the total ISM content. In fact, insight into the "missing" or dark molecular gas (related to the *Formation of Molecular Clouds* above) may come from gamma-ray studies in conjunction with arc-minute studies of atomic hydrogen (HI) and carbon in various chemical states ($C^+$, C, CO) over degree scales.

The major motivation for gamma-ray astronomy has been the origin of CRs, in particular of Galactic CRs. It has long-been thought (since the 1930's) that supernova remnants (SNRs) are the accelerators of such CRs. There are several examples of gamma-ray sources associated with SNRs as cosmic-ray accelerators, based on the spatial match between the gamma-rays and molecular gas, or, characteristic spectral signatures. Such gamma-ray sources tend to be found adjacent to mature SNRs (age $>10^4$ years), which may be indicative of escaped CRs interacting with local molecular clouds. Interpretation of TeV gamma-rays from young SNRs is still unclear however, with indications for both accelerated CRs and electrons. A novel observational method to unambiguously separate CR and electron gamma-ray emission exploits the expected diffusion properties of CRs as they penetrate dense molecular clouds. Only CRs are expected to reach the interiors of dense cloud cores, creating tell-tale X-ray to gamma-ray spectral signatures and morphologies. The diffusion properties of CRs are related to fundamental parameters such as the magnetic field and its turbulence in the ISM. The Figure (from Nicholas et al. 2012) shows the multiply-clumped dense gas CS tracer (colour scale and white contours) overlapping TeV emission (dashed black contours) towards the W28 SNR.

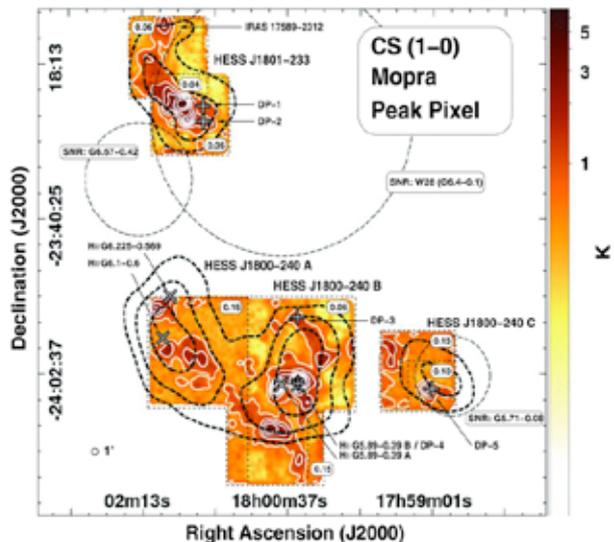

Many Galactic TeV gamma-ray sources are pulsar wind nebulae (PWNe). The TeV emission arises from accelerated electrons via the inverse-Compton processes. Adjacent molecular clouds are responsible for the asymmetric shape of many TeV PWNe. Studies of the ISM are hence important in understanding the development of gamma-ray PWNe. Interestingly, one might also probe for CRs from pulsar winds, which are theoretically expected but only indirectly seen. This can be performed by looking for gamma-rays towards dense parts of the adjacent molecular cloud, and exploiting the CR diffusion aspect as outlined above.

The major mystery from gamma-ray surveys is that >30% of the Galactic gamma-ray sources (at GeV and TeV energies) remain unidentified in terms of the nature of the parent particles (cosmic-ray or electrons), and the counterpart accelerator. Although there are often hints as to potential counterparts, the angular resolution of HESS (~10 arc-minutes – the best in >MeV gamma-ray astronomy) and of current large-scale southern CO surveys (~4–8 arc-minutes) that it is compared to represents a major difficulty in identifying sources. Many of the unidentified sources are thought to be evolved SNRs or PWNe. Better HESS resolution is possible via advanced methods but this is only possible on a few bright sources given limited statistics.

Looking towards the future, the southern installation of the Cherenkov Telescope Array (CTA) will have a TeV gamma-ray sensitivity 10 times better than HESS, and reach arc-minute angular resolution. CTA will therefore provide gamma-ray maps on scales similar to the cores of molecular clouds. Additionally, CTA may detect the Galactic diffuse gamma-ray emission, or certainly, enhanced diffuse emission within several degrees of CR accelerators. Issues such as the distribution and properties of the ISM, B-field turbulence and alignment, plus accelerator age and distance to the target ISM all play a role in determining the gamma-ray morphology. Disentangling diffuse gamma-ray components and identifying gamma-rays sources will certainly require ISM studies on corresponding arc-minute scales over the southern Galactic plane.

*Requirements for ISM Studies*
To advance the studies outlined above, we identify below key observational tasks for several Australian telescopes:

- Arc-minute follow-up mapping of dense ($10^4$ cm$^3$) gas towards TeV gamma-ray sources (e.g. CS, NH$_3$) [Mopra 7mm/3mm].
- Arc-minute surveys of low-to-medium dense gas (CO, $^{13}$CO, C$^{18}$O) covering the southern Galactic plane (ideally $|b|<1.5°$) [Mopra 3mm via extension to the current Mopra CO survey].
- Mapping of shocked and atomic gas (1720 MHz OH, HI) to probe for the disrupted ISM [ASKAP, via the GASKAP project, and Parkes SPLASH].
- Characterising the magnetic field and its turbulence in our Galaxy [ASKAP, via the POSSUM project].
- Using ionic species to trace signatures of CR ionisation arising from <GeV CRs [DCO$^+$/HCO$^+$; possibly with Mopra 3mm].

# High-mass star and cluster formation

Key question: *How is the mass assembled from a dense clump onto stars within a cluster?*

*Overview*
High-mass stars (M> $8M_\odot$) form deeply embedded within dense molecular clumps, typically surrounded by a cluster of lower-mass stars. Thus, understanding how high-mass stars form is directly linked to how clusters are formed. High-mass stars are rare, evolve quickly, and have a profound impact on their immediate surroundings. As such, it has been difficult to observe the detailed processes by which the material within a clump is assembled onto an individual high-mass star and its associated cluster of lower mass stars. Tracing this mass assembly is incredibly important for distinguishing between cluster formation scenarios.

*Identifying and characterising cluster-forming clumps*
In order to understand the process by which clusters form, we require a large sample of dense clumps that are well characterised and that span a broad range in their star formation activity. Significant progress in identifying cluster-forming clumps has been made in the last few years with the combination of Galactic plane surveys of the IR and dust continuum emission (GLIMPSE, MIPSGAL, Bolocam, ATLASGAL). Targeted follow-up surveys of the dense gas, such as MALT90, not only provide distances, but also valuable information about the kinematics and chemistry within these clumps. Combined, these datasets allow us to select out clumps in all stages of their evolution, from those that are quiescent and starless, to those that already harbour a high-mass star and cluster.

*Kinematics, chemistry, and physical conditions on large scales: SuperMALT*
MALT90 (see http://malt90.bu.edu) observed over 2,600 high-mass star-forming clumps within the Galaxy. Using the Mopra telescope, it obtained small maps in 16 molecular lines near 90 GHz around each of the ATLASGAL-selected clumps. The low-J transitions of the molecules included in the MALT90 survey traced a combination of the cold, dense and hot/shocked gas. While these data reveal a wealth of information about the gas morphology, chemistry, and kinematics, alone, they are unable to constrain the physical properties of the clumps. In order to characterize well the various stages in the evolution of a cluster-forming clump, a new, large APEX program – SuperMALT – is about to commence. It will observe the higher J transitions of these molecular species toward a carefully selected sub-sample of the MALT90 clumps. The combination of the Mopra and APEX data will constrain the molecular excitation and allow the determination of accurate temperatures and column/volume densities within each clump. Reliably determining these parameters is critical to establishing their evolutionary stage. With an accurate measurement of their global properties, the best candidates will then be selected for detailed follow-up with ALMA with the goal of testing star and cluster formation scenarios.

*Fragmentation and kinematics on small scales: testing cluster formation scenarios with ALMA*
While these recent Galactic plane surveys have identified a large number of dense, massive clumps where the next generation of cluster formation will likely occur, progress in understanding the process by which clusters form has been limited primarily due to the lack of both angular resolution and sensitivity of the observations. In order to test cluster formation scenarios, observations are needed of the small-scale material within clumps that have sufficient mass to form a cluster but show no evidence for active star formation.

Only with observations of the natal gas and dust in a cluster-forming clump well *before* the onset of star formation can we begin to understand the process by which the material is transferred from the large reservoir within a clump onto protostellar cores and ultimately the stars. To make progress we need the ability to:

- image the cold, dense natal gas *and* dust on very small spatial scales (<0.1pc),
- detect both low- and high-mass protostellar cores (~few solar masses),
- cover a combination of optically thick and thin molecular species simultaneously,
- measure the motion of the cores within a clump and with respect to each other.

Fortunately, the revolutionary capabilities of ALMA will provide the necessary observations, making it possible to observe the small-scale processes and, for the first time, test cluster formation scenarios.

Capabilities need for progress
- Strong international collaboration. To make progress and to address "big questions" requires a large team with a broad range in expertise but that is coordinated to achieve a common goal.
- Access to Mopra, APEX, and ALMA.
- Direct connection to theorists. Currently there are two competing theories describing cluster formation with very different predictions for the initial conditions within a cluster-forming clump. To distinguish between them we need both the observations and techniques to turn the simulations into observable quantities that we can compare to data. Thus, a strong collaboration between observers and theorist is critical.

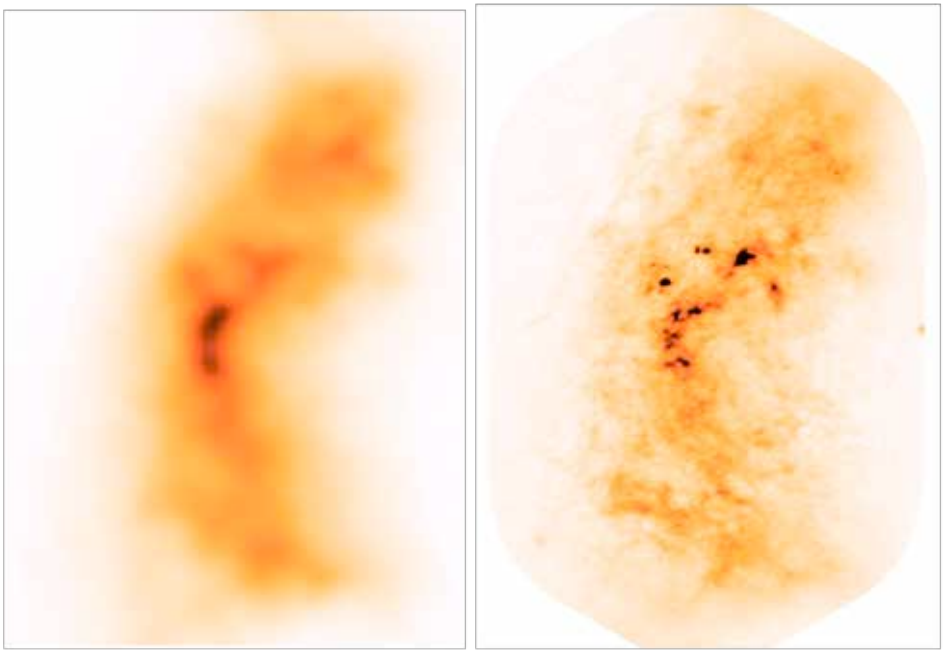

*An example of a cluster-forming clump (left: JCMT image of the 450μm dust continuum emission, 7.5 arc-second angular resolution) and the significant improvement in tracing the protostellar cores provided by ALMA (right: ALMA image of the 3mm dust continuum, 1.7 arc-second angular resolution). Future observations like these will provide much needed information about the formation and early evolution of clusters make it possible to test, for the first time, cluster formation scenarios. Credit: ALMA observatory.*

# Charting the Dense Molecular Galaxy

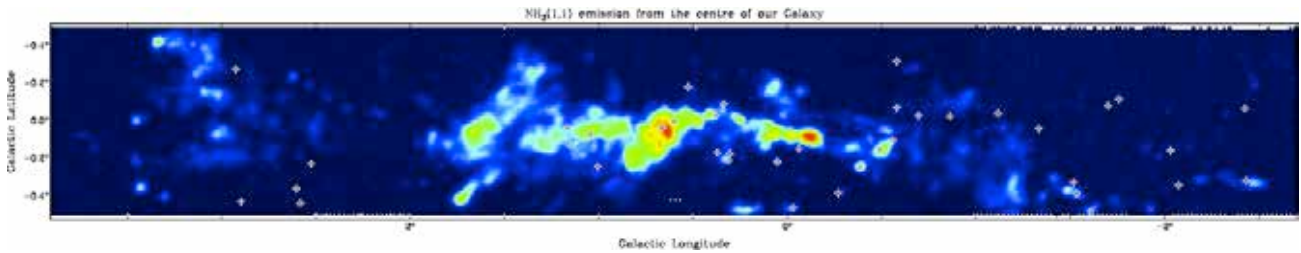

*Ammonia overlaid with the position water masers across the central 5 degrees of the Galaxy, as seen by the Mopra HOPS survey at 24 GHz. See www.hops.org.au*

Understanding the evolution of galaxies remains one of the major challenges for astrophysics. Galaxy evolution is intimately linked to star formation, as the star formation process injects enough energy and momentum to transform the interstellar medium (ISM). Furthermore, star formation enriches the ISM with heavy elements and determines the structure of galaxies. However, the scale on which star formation directly acts with the ISM is typically too small to be studied in detail in external galaxies. Our best option for understanding the interaction between star formation and evolution on a galaxy-wide scale is to study our own Milky Way Galaxy, where modern radio telescopes can resolve such interactions: *galaxy evolution begins at home*.

The dense molecular hydrogen clouds where stars form are hard to observe directly; we rely on a series of tracers of molecular hydrogen, such as dust continuum emission and molecular line emission.

Whilst the continuum view of our Galaxy is now well known, the view though molecules is not so well understood. However, molecule observations offer much more useful information than can be obtained through continuum measurements: we can study kinematics/dynamics of the Galaxy, as well as its 3D structure. Molecules allow us to measure physical quantities such as density, temperature, turbulence, cosmic ray ionization and star formation capacity. In addition, through molecules we can study interstellar chemistry and the formation of more complex molecules on the road to the natural evolution of life.

The most commonly used molecular line tracer is CO (see the *Formation of Molecular Clouds* program discussed earlier). CO has the advantages of being easily excited in the molecular ISM, so that it traces most molecular gas, and having easily observed optically-thin isotopologues. Its disadvantages are that it does not discriminate well between densities in molecular gas, and it freezes out at very low temperatures. These disadvantages make it an imperfect tracer of the dense cold clouds where the first stages of star formation happen. Better information on these clouds can be obtained by dense gas tracer molecules that require higher densities for thermal excitation, such as ammonia ($NH_3$) and cyanoacetylene ($HC_3N$). These dense gas tracers may be observed in the 20-28 GHz band, compared to the 115 GHz needed for CO. This in turn relaxes the requirements for instrumentation and quality of site.

Untargeted CO surveys of our Galaxy have been carried out, and are being continually improved. But nearly all previous molecular line work has been carried out towards targeted sites of star formation. This incurs a necessary limitation that we cannot probe the full population of molecular objects in the Galaxy. Only through untargeted surveys can we learn about the full range of conditions that give rise to molecular line emission.

The small number of previous untargeted surveys of the Galactic plane in dense gas tracers have been highly productive in helping us understand the interstellar medium and the star formation process, but has also shed light on previously unknown phenomena, such as new maser species (Voronkov et al. 2011; Walsh et al. 2007), high velocity maser components associated with jets (Walsh et al. 2009), the discovery of possibly the only super star cluster precursor in the Galaxy (Longmore et al. 2012) and the quantification of the star formation deficit in the centre of the Galaxy (Longmore et al. 2013). These are all programs conducted with Australian radio telescopes: Mopra, Parkes and the ATCA. Thus, the value in future untargeted surveys lies both in understanding the known and discovering the unknown.

HOPS (the $H_2O$ southern Galactic Plane Survey; see www.hops.org.au) was carried out at Mopra, covering the 20-28 GHz band principally for the dense gas tracers of ammonia and water masers. It is limited by its large beam size, making it only sensitive only to Giant Molecular Clouds across the Galaxy. With a more sensitive survey, one able to detect typical star-forming clouds (~$100M_\odot$) across the entire Galaxy, a new field of star formation and interstellar medium research is possible. With such a sensitive survey, it would be possible to conduct a complete census of dense clouds that produce all the high mass stars in the Galaxy. This will help address an important question in high mass star formation: do such stars only form in high mass clusters? In addition, a survey sensitive to dense gas across the Galaxy will produce the most accurate and clear picture yet of the structure of our Galaxy.

The survey goal requires a larger antenna, with a larger collecting area and less beam dilution of sources – Parkes or the Tidbinbilla 70m antenna. The smaller beams which mean that a survey will take a prohibitively long time to complete. This is assuming that current technology is not advanced. However, we are currently seeing great progress in the development of phased array feeds (PAFs) for the ASKAP telescope. With similar technology at higher frequencies on a dish like Parkes or the 70m, such a survey becomes an attainable goal. With a 100-element feed it would be possible to survey a region comparable to HOPS (100 square degrees) with three times the spatial resolution and twenty times the sensitivity in a similar amount of time as HOPS took (i.e. 2000 hours). This is sufficient to detect a typical low mass star-forming cloud on the other side of the Galaxy. Therefore, such a survey would allow us to characterise the full population of dense star forming cores across our Galaxy.

While the survey described above is the next major step in our understanding of the dense ISM in the Milky Way, incremental progress is also possible. While the 70m has comparatively little time available, technical improvements can make it far more efficient. On-the-fly mapping and a wide band (2 GHz) backend can be implemented. Such enhancements would offer the prospect of obtaining sub-square-degree maps of multiple transitions. This includes the higher ammonia transitions [i.e. (3,3) & (4,4) in addition to the current (1,1) & (2,2) used] to measure the dense core temperatures, as well as the sensitivity to map the linear cyanopolyynes (e.g. $HC_3N$, $HC_5N$) so further probing the physical conditions in such dense gas.

The case for research into the dense molecular ISM can be summarised thus: incremental technical advances at Tidbinbilla (together with existing facilities at Parkes) may be used for targeted surveys of molecular clouds, mainly based on the HOPS results. However, the next major advance in our understanding of dense molecular gas and star formation throughout the Milky Way requires a significant time allocation on a Parkes-class telescope equipped with a multibeam feed – such as a phase array feed (PAF).

# Galaxy Evolution and the Diffuse Atomic Gas – GASKAP and HI

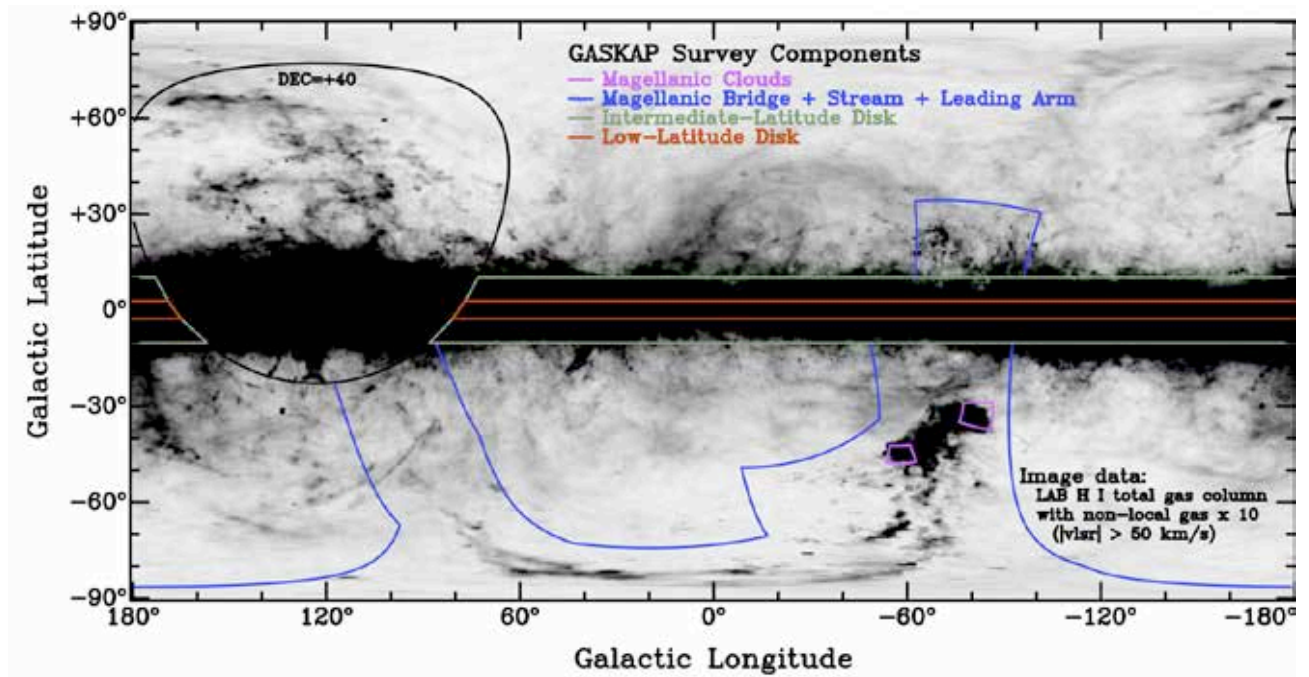

*The GASKAP (Galactic ASKAP) spectral survey areas overlaid on an HI column density image.*

*GASKAP:* The GASKAP survey planned for the ASKAP telescope is a spectroscopic mapping project covering the Milky Way (MW) disk and lower halo and the Magellanic Clouds (MCs), Bridge (MB) and Stream (MS) using the 21-cm line of atomic hydrogen (H I), the 18-cm lines of OH (1612, 1665, and 1667 MHz), and the comb of recombination lines in between. The survey focuses on the generic physical processes that drive galaxy evolution by revealing their astrophysical basis here at $z = 0$. GASKAP will provide a new and vastly improved picture of the distribution and dynamics of gas throughout the disk and halo of both the MW and the MCs.

*Galaxy evolution begins at home.* One of the great challenges of modern astrophysics is understanding how galaxies form and evolve. This is intimately connected with the outstanding problem of star formation: as star formation transforms the interstellar medium (ISM), adding heavy elements and kinetic energy, it determines the structure and evolution of galaxies. While modern cosmological theories can predict the distribution of dark matter in the Universe quite well, predicting the distribution of stars and gas in galaxies is still extremely difficult (Bryan 2007, Tasker et al. 2008, Putman et al. 2009). The reason is the complex and dynamic ISM: simulations reach a bottleneck on size-scales where detailed understanding of star formation, its feedback, and the interaction between galaxy disks and halos need to be included. To make advances in the area of galaxy formation and evolution, we must begin with our home neighborhood where the physics that drives this process can be exposed and studied in detail.

*Studies of the MW and the MCs are essential for determining the physical processes responsible for converting atomic gas to molecular clouds, and ultimately to stars.* What is the relationship between the atomic, molecular and ionized phases of the ISM in different interstellar environments? GASKAP will trace these phases through H I emission, diffuse OH emission and recombination line emission. Comparing these over large areas that contain many kinds of clouds, some forming stars and some not, will show how and where the gas makes the transition from one phase to another. With GASKAP it will finally be possible to obtain images of structures in the HI medium with the richness and detail routinely available in the infrared and optical for the dust and ionized gas.

The motion of the gas in the disk and halo traces both stochastic processes such as turbulence as well as discrete, evolving structures such as chimneys and shells. GASKAP will study this motion primarily in H I cubes that show the velocity structure of the diffuse medium. These are the sources of the feedback that stirs up the gas. GASKAP is designed to trace the effects of this feedback throughout the MW disk and lower halo.

*How do galaxies get their gas?* Cosmological simulations predict that gas accretion onto galaxies is ongoing at $z = 0$. The fresh gas is expected to provide fuel for star formation in galaxy disks (Maller & Bullock 2004). Some of the HI we see in the halo of the MW comes from satellite galaxies, some is former disk material that is raining back down as a galactic fountain, and some may be condensing from the hot halo gas (Putman et al. 2009). However, the detailed physics of gas loss due to stripping and galactic winds, and the fate of the gas that is lost, are still missing in the simulations. How much gas flows in and out of the disk through the halo, how fast does it flow, and what forces act on it along the way? How do halo clouds survive their trip down to the disk? These questions can be studied both through HI structure in the MS, which shows the conditions in the outer halo, and, at low to intermediate latitudes, by the galactic fountain that constantly circulates HI between the disk and the lower halo as seen in chimneys and HI high velocity clouds.

*GASKAP will trace phase changes in the gas on its way to star formation.* In HI absorption, GASKAP will be an even greater advance over existing surveys than it is in HI emission, with astrophysical results to match. The HI absorption spectra GASKAP produces will yield a rich set of gas temperature, column density, and velocity measurements over most of the Galaxy. Anchored to these will be complementary, contiguous maps of the cold HI structure and distribution from HI self-absorption (HISA) against Galactic HI background emission (Gibson et al. 2000, 2007). HISA arises from $H_2$ clouds as well as dense HI clouds actively forming $H_2$, so it directly probes molecular condensation prior to star formation (Kavars et al. 2005, Klaassen et al. 2005). In fact, a cloud's age can be measured by comparing its HISA and molecular content to appropriate models (Goldsmith & Li 2005, Goldsmith et al. 2007). GASKAP's low-latitude survey will easily map the HISA content of 10–20 K clouds with $N_{HI} >$ a few x $10^{18}$ cm$^{-2}$. This sensitivity – enough to see the ∼1% trace HI in $H_2$ cores, plus warmer gas in HI envelopes – will be applied to most of the Galactic disk, enabling comprehensive population studies of $H_2$-forming clouds, including their proximity to spiral shocks (Minter et al. 2001, Gibson et al. 2005a). GASKAP HISA will offer a rich new database for rigorous tests of theoretical models of gas phase evolution in spiral arms (Dobbs & Bonnell 2007, Kim et al. 2008), including phase lags between spiral shocks and star formation (e.g., Tamburro et al. 2008). On much smaller scales, the turbulent froth of HISA filaments that appear to be pure cold H I will be revealed at threefold finer angular and velocity resolution in GASKAP than in prior synthesis surveys, with sufficiently improved sensitivity to follow their spatial power spectrum down to sub-parsec scales where considerable rich structure is already known in isolated cases. This investigation will extend to a wide variety of environments to relate clouds' turbulent support to their stage of molecular condensation. Both 21-cm continuum absorption and self-absorption toward Galactic objects are also helpful for distance determinations (Kolpak et al. 2004). The GASKAP survey is much more efficient in terms of telescope time for this than single pointed observations using the VLA or ATCA.

# Supernova Remnants

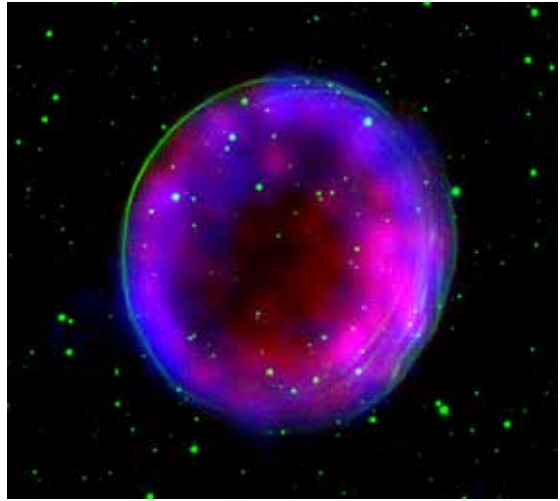

*Colour composite image of Large Magellanic Cloud SNR J0509–6731 made using ATCA radio (red), Chandra X–ray (blue) and Hubble Space Telescope optical (green) observations.*

A supernova (SN) is the energetic "death'" that occurs in certain types of stars, in which the star undergoes a catastrophic explosion, ejecting $10^{51}$ ergs of energy into its surroundings. The explosion is an extremely luminous event which has the potential to outshine an entire galaxy for a brief amount of time. The blast wave from the explosion expels material into the surrounding interstellar medium (ISM) at velocities exceeding 10,000 km/s, sweeping up all gas and dust it encounters. This swept-up material is then heated and goes on to form the forefront of the shockwave and is the basis of the supernova remnant (SNR).

The current belief is that there are two main scenarios which would lead to a star going SN. These are either through what is known as a Type Ia SN – where a carbon-oxygen white dwarf in a binary star system reaches its critical mass of 1.4 $M_\odot$, possibly either through a single degenerate scenario where the white dwarf accretes matter from a larger star or through a double-degenerate scenario, which is the merger of two white dwarfs. The second set of circumstances, known as a core-collapse SN, is thought to occur in much larger stars, greater that 8 $M_\odot$, and starts once this progenitor star begins fusing iron at its core. Due to the net loss in energy in fusion of endothermic elements such as iron, the star quickly collapses in on itself, creating a dense central object and expelling the excess material out into the ISM.

SN explosions are the primary source of heavy elements in the universe, and therefore play a significant role in the elemental abundances of the next generation of stars and planets. Type Ia SN eject iron rich material while core-collapse explosion contribute vast amounts of α-elements. As the SN shockwave propagates through the surrounding environment, it shapes the ISM, compressing and heating ambient gas and dust, creating the basis for future star formation. In addition to altering the chemical make-up of galaxies, SN also affect the structure of the ISM and are the missing link to the life-cycle of stars. As the blast wave from the SN pushes through the ambient medium, it compresses vast amounts of gas, which may be responsible for triggering the next generation of star formation in the region.

<u>Big Questions in SNR research</u>
*Effects of ISM on SNR evolution*: By observing and analysing more SNRs and their surrounding environment, we will be able to better understand how environmental factors can alter the evolution of a SNR, from its morphology to the magnetic field of the remnant.

*Σ–D relation*: This relationship may provide a better understanding of the environment and the intrinsic properties of the SN explosion and progenitor. This is a controversial topic; a larger sample of SNRs at known distances would be beneficial in the application of this relation.

*SNR progenitor systems*: It is still unclear exactly what types of progenitors or, in some cases, progenitor systems are the basis of a SN explosion. The determination of companion stars in single degenerate Type Ia explosion is needed to see whether such SNRs originate from a double degenerate scenario (as might be expected if the companion star were over luminous), or whether they originate in single degenerate systems, where the companion star was a "supersoft" source.

*Mixed morphology or Type Ia SNRs*: This is a sub-class of remnants that to be examined in higher detail in order to understand the role of environments in their production, and/or whether or this arises from a different type of SN explosion altogether, such as a "prompt" Type Ia event.

*Variations of radio spectral index*: The spectral index of a SNR is affected by various factors such as the surrounding density, presence of a central pulsar producing a pulsar wind nebula and also age. With a more in-depth study of the change in the spectral energy distribution over time, we could better understand the underlying processes involved.

There is no doubt that SNRs play an important role in driving the evolution of the interstellar medium of a galaxy. Besides the Magellanic Clouds, SNR surveys in other galaxies are limited by poor resolution and sensitivity of X–ray and radio instruments and hence can only be carried out in the optical using the SII/Hα ratio. Of the SNR candidates identified at a distance of a few Mpc, ~50% are unresolved, ~20% are larger than 100pc, and ~50% have only modest SII/Hα ratios. These properties suggest that many of the candidates are either not SNRs or are unusual remnants.

In order to address above mentioned *questions* we ultimately rely on detection and analysis of a larger sample of SNRs, as well as higher resolution studies. This will be possible given the next generation of telescopes, from the *eRosita* X-ray telescope to the *James Webb Space Telescope* and the myriad of new radio instruments, including the Square Kilometre Array (SKA), the Australian Square Kilometre Array Pathfinder (ASKAP), the Murchison Widefield Array (MWA) and the Atacama Large Millimeter/submillimeter Array (ALMA). The ability of these new telescopes to survey the sky at a much faster rate, while also increasing resolution and sensitivity, will enable an in depth examination of the nature, composition and environment of SNRs. Australian involvement in these new generations of telescopes predominately includes projects such as the ASKAP, MWA and the SKA.

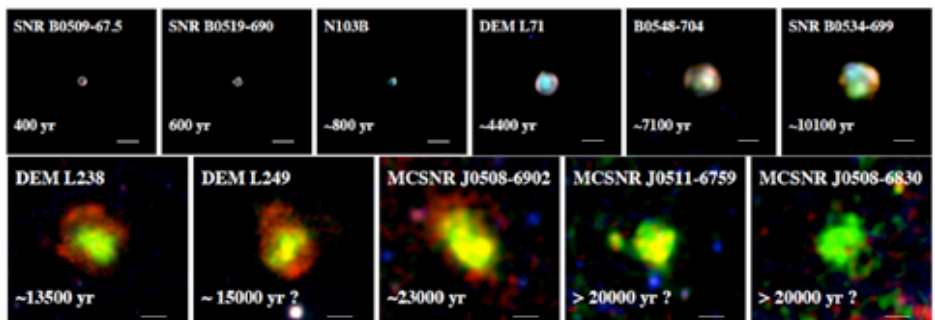

*Sequence of all Type Ia SNRs known in the LMC obtained from Chandra or XMM-Newton observations, shown on the same scale (the white bar is 1 arcmin), and sorted by increasing age. Chandra data were used for the three smallest SNRs. In the (other) XMM-Newton images the red, green and blue components are soft (0.3-0.7 keV), medium (0.7-1.1 keV) and hard (1.1-4.2 keV) X–rays. The medium band is dominated by Fe L-shell lines, and the iron-rich interior, appearing greenish, is readily distinguished from the softer, fainter shells of the more evolved Type Ia SNRs (second row). Image and caption from Maggi et al. (in prep).*

# The Role of Magnetism and Turbulence in the Galactic Ecology

Traditionally the magnetic field of the Galaxy has been separated into large-scale 'ordered' and small-scale 'turbulent' components. Both fields follow the gas on different scales and must be part of any clear picture of the Galactic ecology.

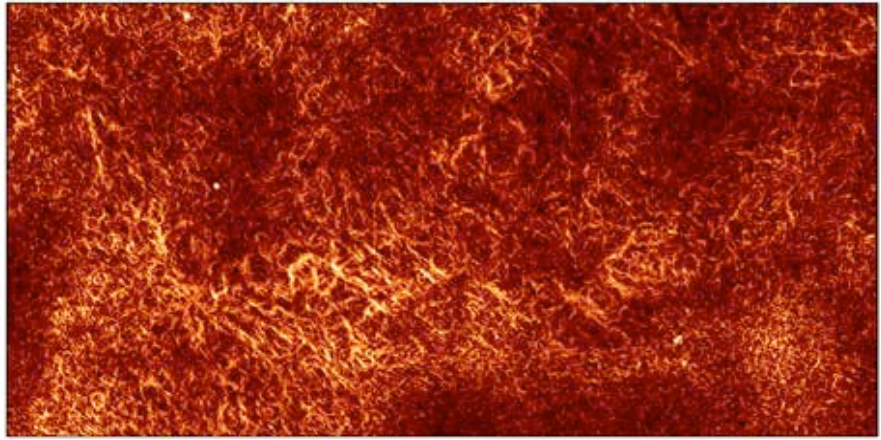

*Image of turbulence in a magneto-ionic medium (Gaensler et al 2011). The spatial gradient of the polarised intensity vector illustrates filaments where the magnetic field or the electron density change within a beam.*

*Large Scale Magnetic Field*

The ordered field has been detected on scales of kpc and is strongest (1-2μG) in the disk of the Galaxy, but also extends into the thick-disk / halo. We still do not understand in detail the processes which generate coherent Galactic magnetic fields. The disk field can be explained using the classical mean-field dynamo theory, i.e., amplification of existing 'seed' fields by small scale turbulent and rotational motions ($\sim 10^9$ years, Wilkin et al. 2007). Dynamo theory can produce a number of different disk field morphologies, with the simplest being a quadrupolar field, of symmetric sign above and below the mid-plane. By contrast, galactic halos are predicted to exhibit dipolar fields of opposite sign above and below the disk.

Much work has already been done to investigate the ordered field of the Milky Way. Observational studies have used the Faraday rotation effect to measure the line-of-sight magnetic field strength towards polarised background AGN Galactic pulsars. The 3-dimensional field can be modelled combining rotation measures (RMs) and pulsar dispersion measures (DMs) with an estimate of the electron-density. Models of the disk field have been produced, for example, by Sun et. al. (2008) and Jaffe et. al. (2010) using data from the SGPS and CGPS surveys. Unfortunately, the sparse spatial sampling of current rotation measure catalogues cannot distinguish well between multiple disk-field geometries. The Sun's location within the Milky Way adds to the difficulty as local magnetic features dominate RM maps and obscure the signature of the large scale field. Most recently, observations on the Very Large Array by Mao et al. (2012) of extragalactic point source RMs support the even parity of the disk field. A more tightly sampled RM-grid is needed to progress understanding of the Galactic magnetic field, on scales of less than an arcminute.

*The Turbulent Magnetic Field*

The turbulent magnetic field has a magnitude of 3-5μG and is thought to be more isotropically distributed than the ordered field. Turbulence has the singular ability to transfer energy from kpc to 1000-km scales, altering the structure and proportion of gas in different ISM phases. Large-scale turbulence moderates the formation of filaments in the diffuse ISM ($\sim 0.1 cm^{-3}$, $T\sim 10^4 K$), and can lead to the subsequent formation of giant molecular clouds (GMCs, $n\sim 100 cm^{-3}$, $T\sim 10K$) in colliding gas flows as HI gas is compressed and cools. At the same time, turbulence and magnetic fields act to suppress star-formation within the clumpy GMCs by supporting these clumps against self-gravity, limiting the feedback of energy into the ISM. Supersonic turbulence dissipates on short timescales, requiring constant driving to maintain observed levels, yet we do not have a good picture on what scales power is injected into the ISM, i.e., what are the dominant driving sources?

Our lack of knowledge is, in part, because the fundamental parameters of interstellar turbulence (e.g., the sonic Mach number) are difficult to determine. Current observations lack the sensitivity and resolution to image the small-scale structure associated with turbulent motion. Past work has also focused on interpreting the distribution of gas column density, from which chemical and physical effects need to be disentangled.

An exciting advance was made by Gaensler et al. (2011), who directly imaged interstellar turbulence via linearly polarised radio emission in the Galactic disk. Calculating the magnitude of the spatial gradient of polarised intensity reveals a network of tangled filaments corresponding to boundaries across which the electron density or line of sight magnetic field jump within a small spatial scale. These discontinuities do not correspond to any features in Stokes I, HI or Hα maps and are likely direct result of magnetohydrodynamic turbulence in the diffuse, ionised ISM. By analysing the moments of the probability distribution function and the filament topology we can constrain sonic Mach number (see Burkhart et al. 2012). Application of such techniques to new broad-band radio data will determine the turbulent properties as a function of longitude and latitude.

*Facilities and Surveys*
Australian astronomers are in a strong position to advance research on Galactic magnetic fields and turbulence. Local facilities such as ASKAP and the MWA will measure previously untapped parameter-space and image the sparsely-mapped southern sky. ASKAP will deliver an new view of the ISM in polarised light and atomic gas. The POSSUM project aims to observe extragalactic rotation measures spaced by ~30″ over the whole sky. Magnetic fields in discrete Galactic objects (such as HII regions, SNRs and wind-blown bubbles) impose a signature on the RMs, and their turbulent properties may be determined from the scatter. Away from the mid-plane the absolute value and scatter of RMs will feed directly into Galactic dynamo models.

The wide-band capabilities of ATCA directly complement ASKAP and allow imaging of turbulence in the warm ionized medium, resolving the turbulent filaments and measuring fundamental properties such as the sonic Mach number. Wide band (1-9 GHz) spectra are also crucial to understand the physics of the background sources and removing any systematic effects.

The MWA is already producing maps of the southern sky from 80-300 MHz. Radiation is easily depolarised at long wavelengths, so towards the Galactic plane the MWA preferentially images local polarised structures. Cubes of the polarised mission as a function of frequency will reveal the detailed magnetic geometry of the Fan Region, the North Polar Spur and the Local Bubble. Away from the mid-plane the MWA is sensitive to diffuse emission, which traces the extended field of the halo and weak local fields. RM synthesis over the full bandwidth will result in an unprecedented resolution of 0.3rad m$^{-2}$ allowing detailed study of turbulence and magnetism in old SNRs.

Current international surveys are pointing the way for ASKAP. GALAFACTS covers 1225–1525 MHz on Arecibo, and delivering an RM-grid ~5 times less dense than ASKAP will. GMIMS has also just finished mapping the northern and southern sky from 300–1,800 MHz.

*Tools*
The sheer size of these datasets makes analysis difficult. It is essential that Australia invests in the tools necessary to query and slice multi-wavelength astronomical data. By necessity, researchers will need to use linked virtual observatory tools to visualise complex *n*-dimensional data and perform robust statistical analysis. Some of these tools already exist (see Goodman et al. 2012 for a review) or are in the planning stage (e.g., the CASDA archive).

# The Environments for Complex Organic Molecules in Space

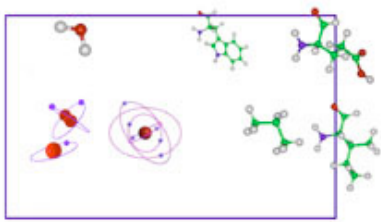 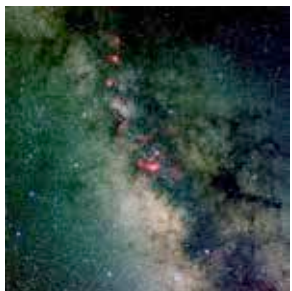 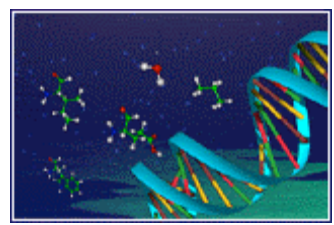

*Complex organic molecules and the southern Galactic plane*

Few questions challenge the imagination more than the question of whether there is life in the Universe, other than the Earth.  Could there be, or was there once, life elsewhere in the Solar System?  Does life also exist elsewhere in the Galaxy?  If so, could life on Earth have been seeded from the interstellar environment?  The environments where complex organic molecules are found in space are therefore natural regions to study in order to address such questions.

We need to know what are these prospective environments and how they have come to be.  A natural starting point for the search is our own Earth – how might one find similar objects, whose surfaces are covered in liquid water and lie within a habitable zone around another star?  However, that may be too confining for the search of parameter space.  As the discovery of liquid methane on Titan shows, liquids other than water may be abundant elsewhere on a planetary surface, and exist within a hydrocarbon environment.  While life may require a fluid medium to sustain the transport of materials to it, and the waste from it, perhaps that medium need not only be water?

These environments on planets and satellites have themselves formed from processes that take place within the dusty disks that surround protostars, molecular cloud cores which have gravitationally collapsed within cold clouds of gas.  The extended surface areas of the disks and clouds make their detection and dissection easier than that of the planets which they ultimately produce.  The clouds themselves are also rich in molecules, with a complex organic chemistry underway – taking place on the surfaces of dust grains and in the gas phase as the protostar heats up and evaporates material from the grain mantles.  Do biogenic molecules exist inside these clouds?  Species such as acetic acid, glycolaldehyde (a sugar), ethylene glycol (anti-freeze) have been found in some sources, and even glycine (the simplest amino acid) found in a few meteorites.  However, it has yet to be detected in the interstellar medium.  Do chiral molecules exist and if so what is the balance between right-handed and left-handed versions of the molecules?  Can these molecules then seed planets with the organic material needed for the precursors to life?  If so, are the origins of life similar across the Galaxy?

These questions may all be addressed through precise observations at infrared, millimetre and centimetre wavelengths.  They demand high spatial resolution, to resolve structures on scales of a few arc-seconds to milli-arcseconds, high dynamic range to simultaneously measure faint objects adjacent to bright stars, and large collecting areas to provide the sensitivity to discern the necessarily weak emission signatures.

Protoplanetary disks will emit radiation that may be detected from thermal infrared wavelengths ($\lambda > 3\mu m$) to centimetre bands ($\lambda \sim 1cm$), depending on the temperature and size of the regions and the column density of the emitting material.  With angular extents of up to several arc-seconds for objects within 100pc of the Sun, millimetre and sub-millimetre interferometers with kilometre-sized baselines will be capable of resolving the extent of cold (T < 100K) disks.  Such observations will provide the first indications of the presence of incipient protoplanetary disks within star forming cores. In the warm central regions of solar system-sized disks, mid-infrared observations with current generation telescopes will be able to detect the emission from arc-second-sized regions, but are insufficient to resolve more than the most basic structures. Planets themselves will not be discernable, but it is possible that gaps carved out of the disk due to presence of forming planets may be seen. Antarctic infrared telescopes would be particularly potent for such studies, combining the high sensitivity resulting from the low background with the high angular resolution facilitated by the stable atmosphere. Mid-infrared interferometers with at least 30m baselines (or diffraction-limited ELTs of this diameter) will be needed to resolve structure on a 0.1″ scale (e.g. at Jupiter-orbit scale for a system 50pc away). 100m baselines or apertures will be needed to examine structure on the AU-scale. Spectrometers operating across the windows from 8 to 30μm will provide information on the physical characteristics of the grain material, in particular with the ability to distinguish between crystalline and amorphous material, as might be expected if the inner portions of disks undergo substantive heating and processing during the planetary formation phase.

The emission from the inner regions of disks will likely be optically thick at infrared wavelengths, but optically thin in the millimetre and centimetre bands. Observation in the radio is therefore needed to measure the amount of material present. For instance, for a disk of mass $0.1 M_\odot$ and radius 10AU, the emission would be optically thin for $\nu < 10$ GHz. In order to resolve structure in the inner AU of a source 100pc away requires a spatial resolution of 0.01″, equivalent to a 500km baseline at 10 GHz, and also a collecting area of $\sim 10^6 \, m^2$ for the emission to be detectable. This is the realm of the high-frequency component of the SKA. Furthermore, a critical but poorly understood aspect of planetary formation involves the growth of dust grains, in particular how dust grains of size a few millimetres grow to sizes of a few metres, without smashing themselves apart in collisions (when they reach larger sizes gravity can then take over the accretion process). To examine these phenomena observations at wavelengths comparable to the grain size are needed; i.e. in the millimetre and centimetre regimes.

To study the organic chemistry inside a molecular cloud it is necessary to measure the plethora of spectral lines emitted from the molecules across the sub-millimetre and millimetre regimes. In the sub-millimetre, higher rotational levels are evident, arising from warmer (and therefore smaller) regions, closest to the protostellar objects where, presumably, the 'hot core' chemistry is also being driven. The more complex, and therefore rarer, the molecule, the more emission lines are present, a facet of the complexity of the partition function. The lowest energy lines are also shifted towards the low frequency end of the millimetre spectrum the heavier the molecule, and even into the centimetre bands. To be sure of detecting a rare species many lines therefore need to be measured to distinguish it from the 'jungle' of all the species present. Wide correlator bandpasses are essential to achieve this. The ability to observe across the entire mm-band windows simultaneously, with spectral resolutions sufficient to resolve down to thermal line widths, would greatly facilitate the search for biogenic molecules, allowing a complete inventory of the species present in many protostellar environments to be undertaken. Current correlators have around 10,000 channels, but million channel devices would be needed to provide such a capability. They would make it possible increase bandpasses from the current 8 GHz to 30–40 GHz wide, and so allow the entire 100 or 50 GHz band windows to be measured at once. Such investigations can readily be undertaken using single dish millimetre-wave telescopes. In addition, interferometry with ALMA will further help the process of identification, by providing validation, or otherwise, that any candidate lines found all arise from the same region of space, within the beam of the single dish telescope.